# PHYSICAL DESCRIPTION AND MODELLING OF PAPER STRIP FLIGHTS

## LEONARD MÜNCHENBACH AND LEO NEFF

WINNING PROJECT AT BUNDESWETTBEWERB JUGEND FORSCHT 2020 IN PHYSICS, EUCYS 2020 AND ISEF 2021

## ABSTRACT

Paper strips, e.g. confetti, descending to the floor begin to rotate around the horizontal axis after a short time. This slows down the vertical velocity compared to free fall and adds a horizontal velocity component. The frequency of rotation, the angle of fall and the speed of fall seem to depend on the dimensions and mass of the paper strips.

The physics of this flight can be described, i.e. the movement, the translation of the centroid (energy analysis, force analysis, formula for fall speed v(b,h,m), formula for angle of fall α(b,h,m)), the rotation around centroid (formula for frequency f(b,h,m), rotation speed), the flight path with the results from translation and rotation, the visualization of air turbulence and the loss of energy.

## 1. INTRODUCTION TO THE TOPIC AND IDEA FOR THE PROJECT

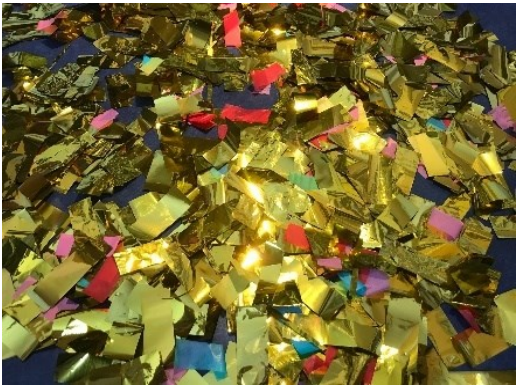

*"But what's the point [...]? And what about applications?... It's like asking mountaineers why they climb mountains. The answer that George Leigh Mallory gave to this question can be summed up as follows [for physicists, too]: he wanted to climb Mount Everest 'because it's there'."* George G. Szpiro [1]

*"There's a lot that physics cannot describe. In my first semester at university, my professor once said: drop a piece of paper. Physics will never be able to describe that."* Bruno Ritter

Confetti cannons are very popular at large-scale events – whether a boxing match, the Champions League final, "Who wants to be a millionaire?" or a concert. Generally glittering or brightly coloured, the confetti attracts a lot of attention. Setting off a confetti cannon in this way provides only a brief source of pleasure – or is it actually surprisingly long-lasting? Small objects fall in the direction of our feet, that is true – but obviously not in free fall, otherwise they would land on the floor very quickly. Our experimental objects are rectangular pieces of paper with a height of between approximately one and 15 centimetres and a width of between approximately three and 20 centimetres. The words "snippet", "shred", "tinsel" or "sheet" [2] could also be used to describe such objects, or they could be referred to in physical terms as "tiles". None of these words offer a perfect match, so we simply decided to call them "(paper) strips".

## 2. QUESTION, STATE OF RESEARCH

When observing a confetti cannon, we wondered whether it would not be possible to comprehend the fall of these paper strips: after all, even though the whole thing appears to be quite random, there is a certain regularity to it as well. Would it be possible to explain this lengthy fall, and potentially even model it? The attractive flutter effect is achieved by firing off countless confetti into the air. We want to keep our experimental set-up as verifiable as possible, however, so we will only observe one paper strip at a time. By carrying out numerous experiments, our aim is to model and predict the trajectory of a paper strip as well as the frequency of the superimposed rotation, and we also want to be able to reproduce the fall based on an energy and force analysis. Based on our research, no one has yet published a research project on this subject, so it is difficult to compare our results with those arrived at by other people.

## 3. FROM THE PRELIMINARY EXPERIMENTS TO THE PROJECT IDEA

In January 2020, we conducted the very first tests using confetti (commercially available confetti, measuring approx. 5 cm x 2 cm). The flight was recorded using an iPad and analysed with Vernier Physics [Fig. 1]. One of the things these preliminary tests showed was that the vertical speed of the paper strip decreases in the course of the fall and settles at about -0.25 m/s. In addition, the frequency of the rotation seemed to be constant. These experiments provided a sufficient basis for a project idea. Our next step was to vary both the dimensions and the grammage[1] of the paper strips in order to be able to generalise the preliminary tests. We then undertook another preliminary experiment in which we captured the flight of the strips with three cameras from the x, y and z perspectives simultaneously in order to be able to experimentally determine the x, y and z position of the strip during flight. To do this, we cut five strips of different sizes from a Hamburg Planetarium flyer (21 cm x 10 cm = ⅓ A4 (original size of the flyer), 21 cm x 5 cm, 10.5 cm x 5 cm, 5.25 cm x 5 cm, 5.25 cm x 2.5 cm, G = 246 g/m²), in which the height and width[2] (and therefore also the mass) were alternately halved. This preliminary test showed if you halve the height, the frequency[3] of rotation increases and the strip flies further. By halving the width, the flight distance decreases and the frequency of rotation hardly changes. The preliminary tests also showed that videos from above and from the front are unsuitable for analysis. A video taken orthogonally to the direction of flight is sufficient to determine the position of the strip as long as the strip flies exactly in a straight line. For this reason, it was clear that a tracking system was in fact superfluous. Another result of these preliminary tests was that in order to achieve reproducibility, the start had to be from a fixed position and a fixed position of the piece of paper. This resulted in the construction of the frame.

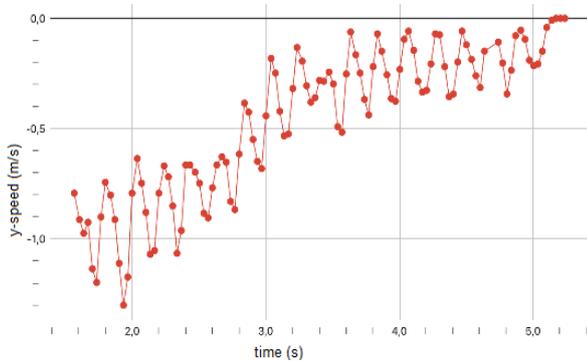

*Figure 1: One corner of the strip was tracked using Vernier Physics*

## 4. EXPERIMENTAL SET-UP

Initially, we held the pieces of paper in our hands before dropping them. To make the experiments more reproducible, we designed a stable frame so that each piece of paper could finally be launched in a virtually identical manner. The frame was built from aluminium profiles by item. [Fig. 2]

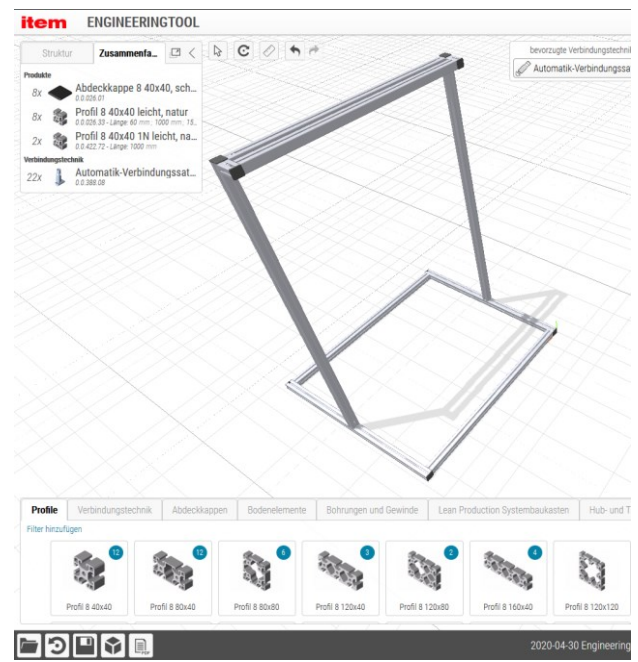

*Figure 2: Profile design*

In order to ensure that the strips started from exactly the same point, we built a mechanism that was able to press a strip against the frame and release it after a few seconds. The CAD programme Solidworks was used to design the components required, along with a holder on the frame, connectors to fit the item profile and butts for different paper formats. These were then 3D printed [Figure 3]. The control system was taken care of by an Arduino run on a programme we wrote ourselves.

In order to idealise the experimental set-up, the paper strips were always to have exactly the same dimensions and be almost perfectly rectangular in shape. In the preliminary tests, we used a paper cutter to cut the strips. For all further experiments, the dimensions of the paper strips were drawn using Inkscape and then cut to millimetre precision with a laser cutter.

---

[1] Grammage: mass per area in g/m²; we use the abbreviation G
[2] From now on we will refer to the longer side as the "width" and the shorter side as the "height".

[3] In the following, we understand frequency to be the reciprocal of the duration of a complete rotation of a strip around the axis of rotation.

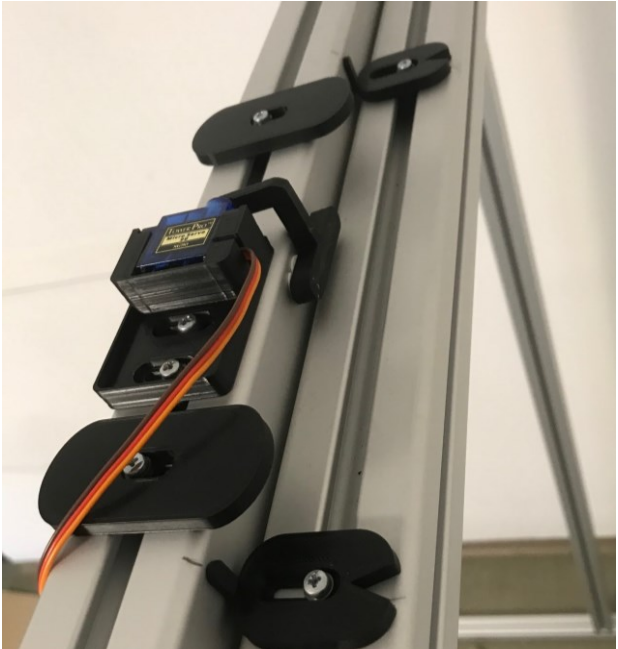

*Figure 5: Attachment parts*

The experimental set-up complete with spatial dimensions was modelled in Geogebra [Figure 3], also in order to clarify the calculation of the spatial position of the strips. The viewpoint B is located at the origin of the coordinate system, the camera K is on the y-axis. M shows the position of any measuring point. The actual positions of the point of the strip T are in a plane parallel to the xz-plane (as long as the flight trajectory is parallel to the xz-plane). This plane intersects the y-axis in S. According to the first intercept theorem, $\frac{\overline{KS}}{\overline{KB}} = \frac{\overline{KT}}{\overline{KM}}$ or $\overline{KT} = \frac{\overline{KS}}{\overline{KB}} \cdot \overline{KM} = q \cdot \overline{KM}$.

By choosing the point of origin in B, T therefore equals
$T(q \cdot M_x, 0.7, q \cdot M_z)$.

The calibration field on the wall is used for dimensioning purposes in Tracker. Another option for dimensioning a track in Tracker is the strip edge. Its length is known and remains the same as long as the strip flies parallel to the xz plane. Both options were used to determine the x and z coordinates of the strip. The y coordinate was discarded because the strip flies parallel to the xz plane. The parallax error in determining the spatial coordinates was compensated for by means of a proportionality factor.

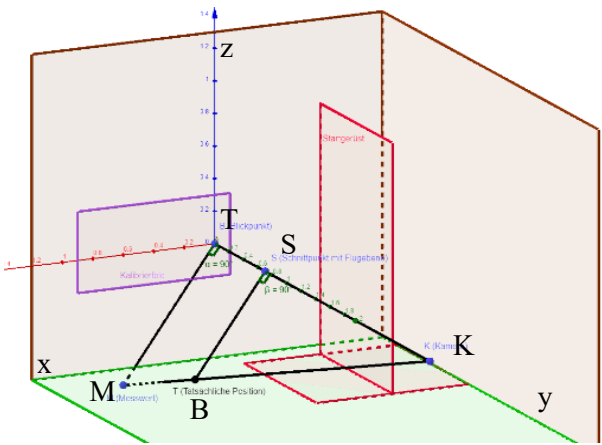

*Figure 4: Experimental set-up*

For our first experiments in spring, we used an iPhone 7. This was attached to a tripod with a self-printed holder [6]. The position of the tripod was chosen to enable ideal filming of the flight. The iPhone 7 is able to record slow-motion videos at 240 fps (8x slowdown) with a resolution of 720 p, which was better than with the iPad used for the preliminary tests. From the summer onwards we were able to record videos at 7680 fps (0.125 s duration, 720 p) or 240 fps (1080 p) using a Huawei P40 Pro. [7][8]

In order to follow the fall of the paper strip exactly, we used the software Tracker [9], which enabled us to view the videos frame by frame. We used the programme in two different ways.

Firstly, it helped us find out the rotation frequency of the paper strips by simply counting the frames. Secondly, Tracker enables the user to place a point in each frame to mark an object. The software is then able to create diagrams using these points and calculate velocities, accelerations, distances, angles, angular velocities, etc. based on any times or coordinates. [Figure 4] The values can also be displayed in a table. What is more, the programme can be used to calculate a centroid between two different points on an image. Tracker also enables analyses to be carried out such as regressions.

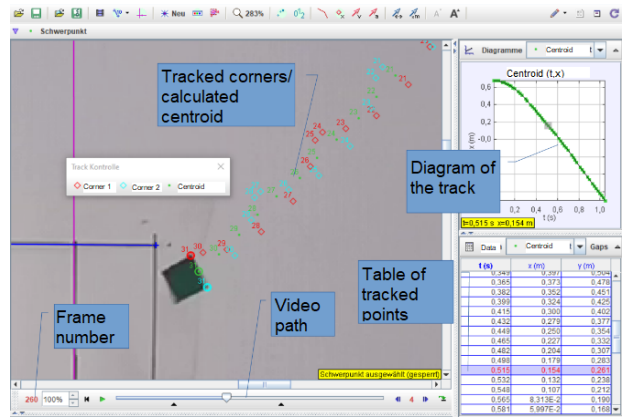

*Figure 3: Tracker*

## 5. EXPERIMENTAL PROCEDURE

### 5.1. THE FALL OF A PAPER STRIP

If we look at the fall of a paper strip, we can divide it into three phases. In the first phase, a strip descends vertically to the floor for a certain period of time. This time span is random and therefore not reproducible. The strip can potentially fall all the way to the floor without changing its position. In the second phase, the lower edge starts to move out of the fall line. This causes the strip to start rotating around the horizontal axis of symmetry. In the process, the axis of symmetry moves away from the line of fall. This phase lasts for between one and three rotations. In the third phase, the strip then rotates at a constant frequency [Fig. 6] and moves towards the floor maintaining an approximately constant fall angle and speed.

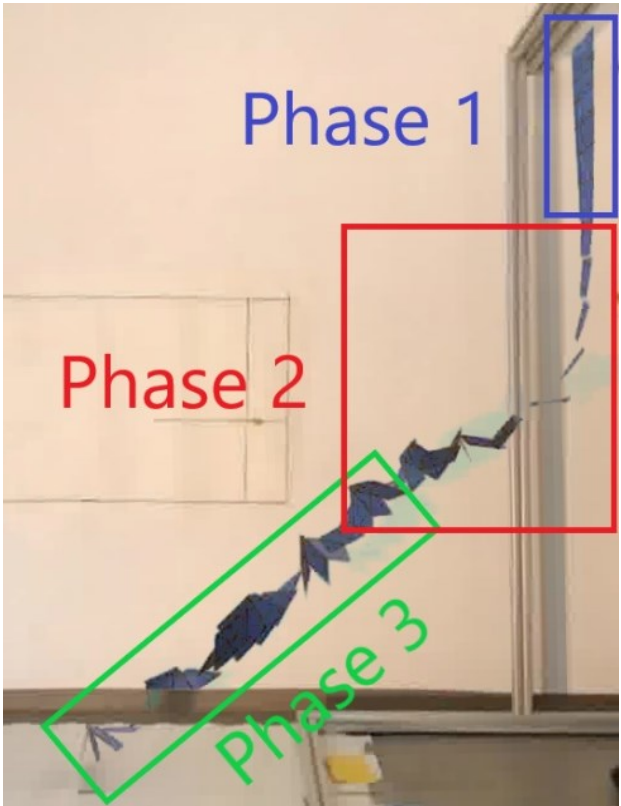

*Figure 6: The three phases of flight*

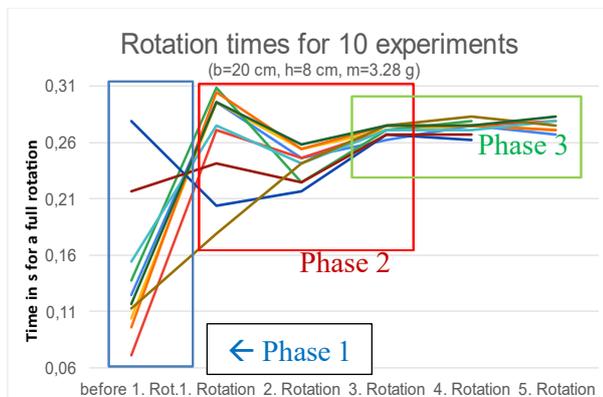

*Figure 8: Scatter of the rotation period across the three phases*

---

[4] We opted to use cm instead of m for better readability

### 5.2. CHANGING THE FREQUENCY BY REDUCING THE HEIGHT OF THE STRIPS

Our initial experiments were dedicated to ensuring the reproducibility of results and testing the experimental set-up. For this we used two geometries (20 cm x 8 cm, 20 cm x 4 cm, G = 205 g/m²)[4].

The confetti pieces fired from confetti cannons measure 5 cm x 2 cm. We wanted to keep the same ratio of width to height and ensure the strip was clearly visible on the video. A height of 8 cm was also practical because it is a power of two and can easily be halved three times. The second strip was half the height, as in the preliminary tests.

A total of 52 videos were recorded for these two measurements, but of these, only ten and sixteen respectively were usable.

We noticed that the time required for a full rotation still varied greatly in the first two rotations. The greater the number of completed rotations, the lower the scatter of the duration of the rotation[5] [Fig. 7] and the frequency. We therefore only considered the last three rotations when carrying out our observations, calculations, modelling and frequency comparisons. The medians, means, quartiles, maxima and minima of the frequencies were also calculated based on the last three rotations of all measurements of a geometry.

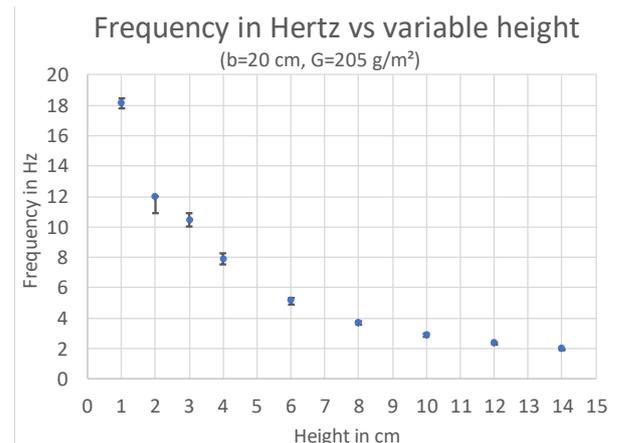

*Figure 7: Frequency as a function of height*

We also found that the period of time before the first rotation differed due to the differing free fall periods. We therefore decided that further analysis of flight distance did not make sense for the time being.

This confirmed our initial assumption that duration of the rotation and frequency depend on the height of the strip. If the height of the strip is halved, the duration of the rotation is halved and the frequency is approximately doubled. In the weeks that followed, we repeated the experiments with additional strip heights h in order to be able to better assess the dependence of the frequency on strip height.

---

[5] Initially we focused on rotation duration rather than frequency

It was confirmed that frequency is roughly doubled when height is halved. [Fig. 8]

It should be noted that the 20 cm x 1 cm and 20 cm x 2 cm paper strips bent in the middle during rotation and therefore exhibited a physically different flight behaviour as compared to the others [see section 6.3]. We considered these strips separately because of their abnormal flight behaviour.

### 5.3. CHANGING FREQUENCY BY VARYING WIDTH, HEIGHT AND MASS

Since we had not initially found width to have any significant influence on frequency, we set about investigating this systematically. Strips with a height of h=6 cm and a width of b=6 cm to 18 cm (G = 205 g/m²) were used for this purpose.

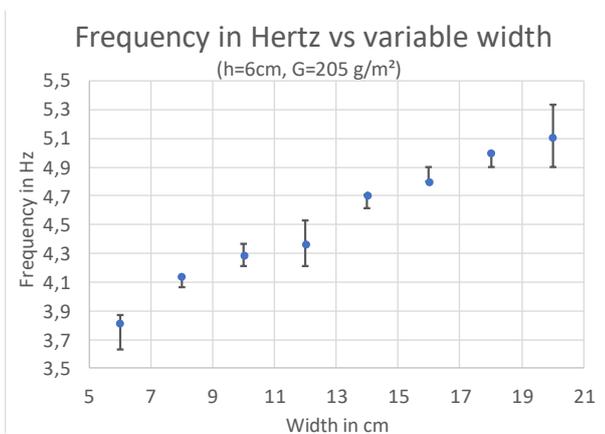

*Figure 9: Frequency at variable width with error bar*

It emerged that an increase in width resulted in an increase in frequency [Fig. 9]. However, an increase in width also meant an increase in mass, which should be taken into account here (see 4.6).

Later, we applied a total of 32 variations of width and height to obtain a better general idea of how frequency depended on width and height. For this purpose, starting from the series of measurements mentioned above (b=20 cm, h=1 cm ... 14 cm or h=6 cm, b=6 cm ... b=20 cm), the width was halved three times and the height was doubled once and halved twice. This resulted in 32 additional b-h combinations. Varying height or width inevitably changes not only the surface area but also the mass. To find out to what extent the mass is relevant to frequency, we carried out 48 further series of tests using grammages of between 40 g/m² and 700 g/m² to find out the influence of mass: same width and height but different grammage, same width and mass but different height, same height and mass but different widths

## 6. EXPERIMENTAL ANALYSIS

### 6.1. TREND FUNCTIONS FOR THE MEASUREMENTS

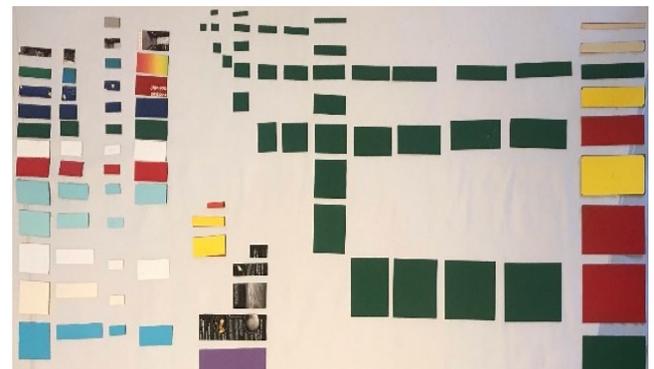

*Figure 10: The paper strips of all the geometries (right) and grammages (left) used*

For all 80 combinations of width, height and mass or grammage [Fig. 10] that we measured, we collated the measurements in individual tables in LibreOffice Calc and Microsoft Excel and calculated the median, mean, minimum, maximum, first quartile and third quartile of the frequency. Finally, we combined the medians, means, minima, maxima, first quartiles and third quartiles of frequency for all 80 combinations in a single table. The spreadsheet programmes allowed trend functions to be established based on the assigned values. This resulted in 12 variations (see next page).

**variable:** h, therefore also m
**constant:** b=2.5 cm, 5 cm, 10 cm, 20 cm; G=205 g/m²
$f_b(h) = c_0 \cdot h^{c_1}$, where $c_0$ and $c_1$ depend on b and also on m

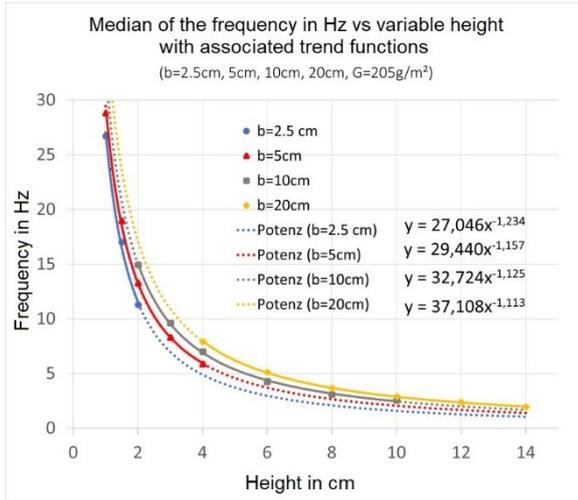

*Diagram 1: Trend line for variable height, R² > 0.995*

**variable:** b, therefore also m
**constant:** h=1.5 cm, 3 cm, 6 cm, 12 cm; G=205 g/m²
$f_h(b) = c_2 \cdot b^{c_3}$, where $c_2$ and $c_3$ depend on h and also on m

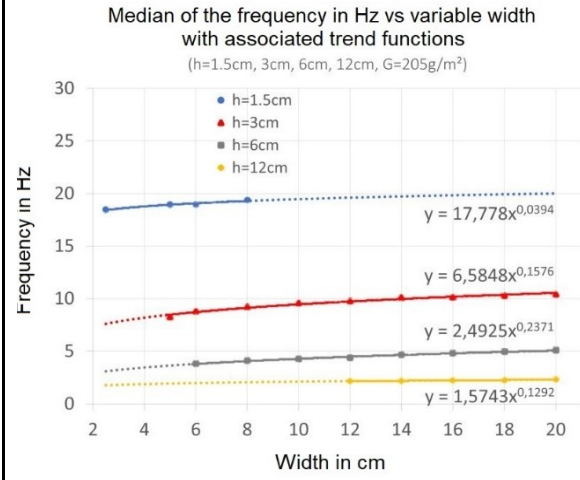

*Diagram 2: Trend line for variable width, R² > 0.94*

**variable:** G, therefore also m
**constant:** b=5 cm, h=2 cm and b=10 cm, h=4 cm
$f(m) \approx c_4 \cdot m^{0.45}$ where $c_4$ depends on b and h, the

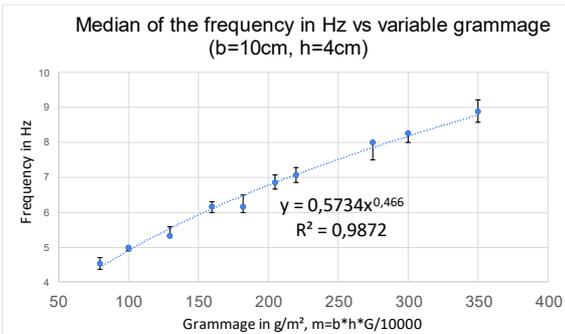

*Diagram 3: Trend line for variable mass*

exponent seems to be approximately constant.

**variable:** G, therefore also h
**constant:** b=10 cm and m=0.6 g
$f_{b,m}(h) \approx 51.074 \cdot h^{-1.555}$

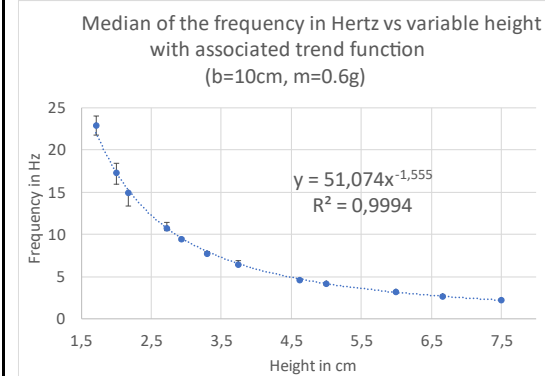

*Diagram 4: Trend line for variable height and constant mass*

**variable:** G, therefore also b
**constant:** b=3 cm and m=0.36 g
$f_{h,m}(b) \approx 12.8949 \cdot b^{-0.245}$

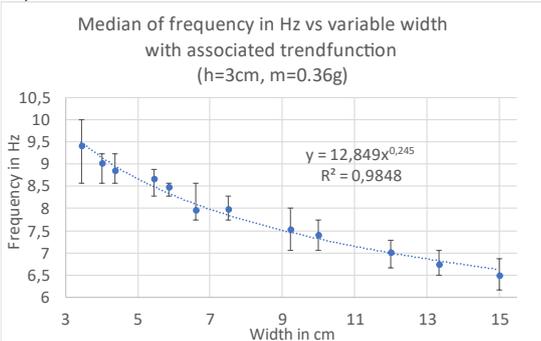

*Diagram 5: Trend line for variable width and constant mass*

For all variations, a power function as a trend function gave the best approximation.

A function of shape
$f(b, h, m) = d_0 \cdot b^{d_1} \cdot h^{d_2} \cdot m^{d_3}$

seemed to us to be a logical consequence of the previous considerations and mathematical modelling of the data measured.

## 6.2. PARAMETER VARIATION

Assuming that the frequency function really is based on shape $f(b, h, m) = d_0 \cdot b^{d_1} \cdot h^{d_2} \cdot m^{d_3}$, we wrote a Python programme that searches for possible values for $d_0$, $d_1$, $d_2$ and $d_3$ so that the formula approximates the measurements as closely as possible.

There are three possible approaches for approximation:
1. Sum of squared absolute errors (`err_quadr`) is minimal
2. Sum of squared percentage errors (`err_faktor`) is minimal
3. Number of function values between minimum and maximum measurements (`counter`) is maximum

For this we used five loops. The outer four loops count through the parameters $d_0$, $d_1$, $d_2$ and $d_3$ at specified intervals in specified increments. The start value and end value of the loop were generously chosen from the function terms of the trend functions. The fifth loop uses width, height, mass, minimum, maximum and median from almost all 80 measured combinations. Combinations in which the strips bent significantly, for example, were discarded. In the fifth loop, an approximation ("naeherung") is calculated for each individual combination. The quadratic error sum is determined from these 80 approximations [Tab. 2].

We also decided that a percentage error might give a different or better approximation, so we also determined the sum of the squared percentage errors. In addition, we wanted to know how many of the approximate values lie between the minimum and maximum of the respective combination.

```
[…]
def formel(c0, c1, c2, c3, cb, ch, cm):
    return c0 * (cb ** c1) * (ch ** c2) * (cm ** c3)
[…]
for d0 in f_range(1, 200, 1):
    for d1 in f_range(-2, -0.001, 0.1):
        for d2 in f_range(-2, -0.001, 0.1):
            for d3 in f_range(0.001, 1, 0.1):
                for key, val in exp_value.items():
[…]
    naeherung = formel(d0, d1, d2, d3, b, h, m)
    err_quadr = err_quadr + (md-(naeherung))**2
    err_faktor = err_faktor + (1-md/naeherung)**2
    if min <= naeherung <= max:
        counter += 1
    […]
counter = 0
exp_err_faktor = 0
exp_err_quadr = 0
[…]
```
*Table 1: Relevant part of the Python programme*

After the first run, it was found that $90 < d_0 < 130$, $-0.3 < d_1 < -0.2$, $-1.6 < d_2 < -1.5$ and $0.4 < d_3 < 0.5$ gave very good approximations. By adjusting the start values, end values and step sizes in the first four loops, we were able to get even closer to optimum values. For $d_0=121.1$, $d_1=-0.258$ $d_2=-1.578$ and $d_3=0.452$, 89% of the approximated values were between the minimum and maximum of the measurements. The percentage deviation from the medians of the measurements was a maximum of 5.84% for all calculated approximated values.

If one assumes that the medians used include measurement errors too, it makes more sense to specify ranges in which the parameters lie. The following therefore applies to the frequency of rotating paper strips:

$f(b, h, m) = d_0 \cdot b^{d_1} \cdot h^{d_2} \cdot m^{d_3}$

with $d_0=121.1\pm15.2$, $d_1=-0.258\pm0.038$, $d_2=-1.578\pm0.030$, $d_3=0.452\pm0.030$, f in Hz;

b, h in cm; m in g; $d_0$ in $\frac{cm^{1.83}}{g^{0.45} \cdot s}$

The exponents of the trend functions (see above) are also found in these areas. As such, the trend functions were simply special cases of the general function we found.

## 6.3. FORECAST – COMPARISON/MEASUREMENT

A function of shape $f(b, h, m) \approx 121.1 \cdot b^{-0.258} \cdot h^{-1.578} \cdot m^{0.452}$ therefore describes the frequency of falling strips very well. In order to test the validity of our modelling beyond the range considered, it was applied to other "strips". Even for "strips" that were far larger and heavier than those used for the formula (e.g., the *Jufo* banners), only minimal deviations resulted. This bears out how accurate the approximation formula is.

| | Measured frequency (median) | Calculated frequency | Error in % |
|---|---|---|---|
| 126 cm x 30 cm (*Jufo* banner 2012, G=608 g/m²) | 1.67 Hz | 1.64 Hz | 1.7% |
| 126 cm x 30 cm (*Jufo* banner from 2013, G=444 g/m²) | 1.98 Hz | 1.92 Hz | 3.2% |
| iTunes card | 8 Hz | 8.41 Hz | 5.1% |
| Original confetti pieces – paper | 4.93 Hz | 4.82 Hz | 2.4% |
| Original confetti pieces –gold | 4.57 Hz | 4.82 Hz | 5.1% |

## 6.4. TRACK OF A FLIGHT

In addition to the rotation frequency, we were also fascinated by the fall curve of the paper strip corners. Determining such a curve is not an end in itself. It can be used to calculate the position of the centroid of a line from two corner points, which in turn is relevant when analysing forces and energy.

The software proved to be a very useful tool in calculating the centroid between two points [see 3.6]. We placed track points on the two corners of the strip that were closer to the camera lens and rotated around the centroid of the edge line during flight. In our case, this is the point on the axis of rotation at the edge of the strip around which the corners (and the horizontal edges) rotate. Tracker then automatically calculates what we refer to as the centroid – the centre of mass.

### 6.4.1. CORNER MOVEMENT

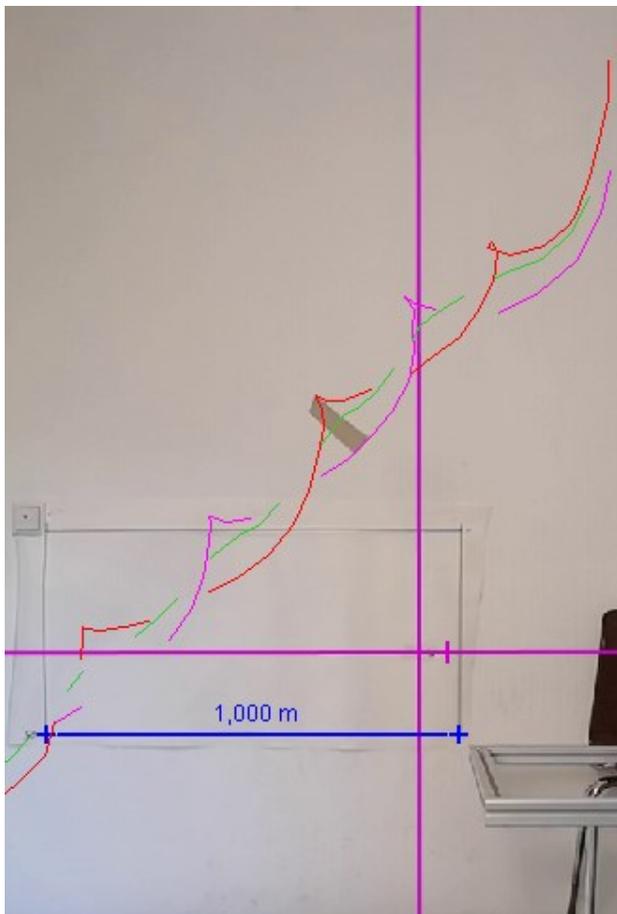

*Figure 12: Track of the two front corners and the centroid of the edge line (green)*

Looking at the diagram of the corner movements in the video, it can be seen that they exhibit periodicity, at least as long as the axis of rotation is parallel to the projection axis [Fig. 11].

The period length is the time a strip takes to complete one full revolution.

The tracking software enables use of the calculated centroid as the origin of the coordinate system. The tracked points are then displayed approximately in circular form around the centroid because the height of the paper strip – i.e. the diameter of the circle – does not change. This also facilitates identification and elimination of tracking errors (see 6.1).

The position of a corner point can therefore be represented in an idealised parametric way on a circle [Fig. 12]:

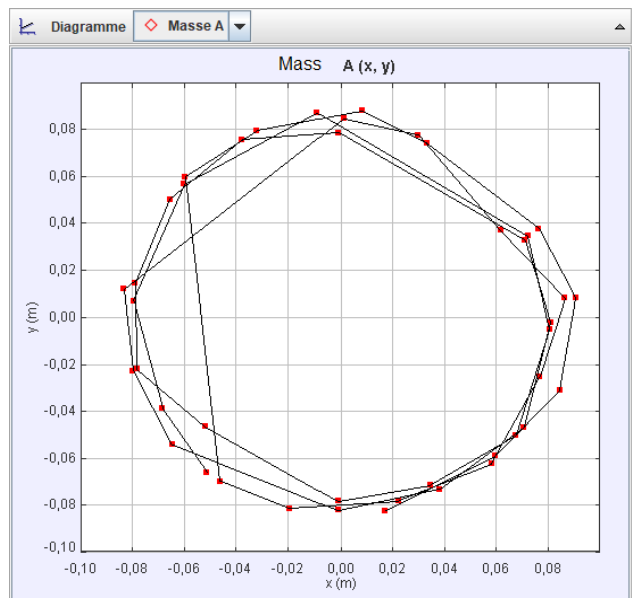

*Figure 11: Movement of the corner around the centroid of the edge line*

(0a) $x(t) = r \cdot sin(2\pi \cdot f \cdot t) = r \cdot sin(\omega \cdot t)$
(0b) $y(t) = r \cdot cos(2\pi \cdot f \cdot t) = r \cdot cos(\omega \cdot t)$
r = radius in m = h/2; f = frequency in Hz; t = time in s; $\omega$ = angular velocity in Hz

The corner points rotate clockwise, so sin and cos are reversed as compared to the definition of the circle.

The centroid does not stay in one place: it moves as the corners rotate around it. They therefore move on a rolling curve (= cycloid).

(1a) $x(t) = r \cdot (2\pi \cdot f \cdot t - sin(2\pi \cdot f \cdot t))$
(1b) $y(t) = r \cdot (1 - cos(2\pi \cdot f \cdot t))$

Tracker allows the velocity $v_x$ and $v_y$ of the centroid to be shown as a function of time.

If we include these velocities in the parameter representation, we arrive at the following:

(2a) $x(t) = r \cdot (2\pi \cdot f \cdot t - sin(2\pi \cdot f \cdot t)) - v_x \cdot t$
(2b) $y(t) = r \cdot (1 - cos(2\pi \cdot f \cdot t)) - v_y \cdot t$

The centroid of the paper strip moves to the left, but the corners rotate clockwise. The rolling direction therefore does not correspond to the direction of movement of the centroid.

The centroid moves approximately on a straight line, but slight fluctuations can also be observed: the period duration of these corresponds to that of the corners. This movement of the centroid can be incorporated as a further correction factor. The movement of the centroid is clearly shown in Tracker.

(3a) $x(t) = r \cdot (2\pi \cdot f \cdot t - sin(2\pi \cdot f \cdot t)) - v_x \cdot t - r_{vx} \cdot cos(2\pi \cdot f \cdot t)$
(3b) $y(t) = r \cdot (1 - cos(2\pi \cdot f \cdot t)) - v_y \cdot t - r_{vy} \cdot sin(2\pi \cdot f \cdot t)$

In summary, it can be said that the movement of a corner – or the axis of rotation – can be described sufficiently accurately by the above formulae with the appropriate r, f, $v_x$, $v_y$, $r_{vx}$ and $r_{vy}$. In order to analyse the forces, the movement of the centroid as a straight line is a sufficient approximation, since the amplitude ($r_{vx}$ and $r_{vy}$) of the fluctuation of the centroid around a straight line is relatively small [Fig. 13].

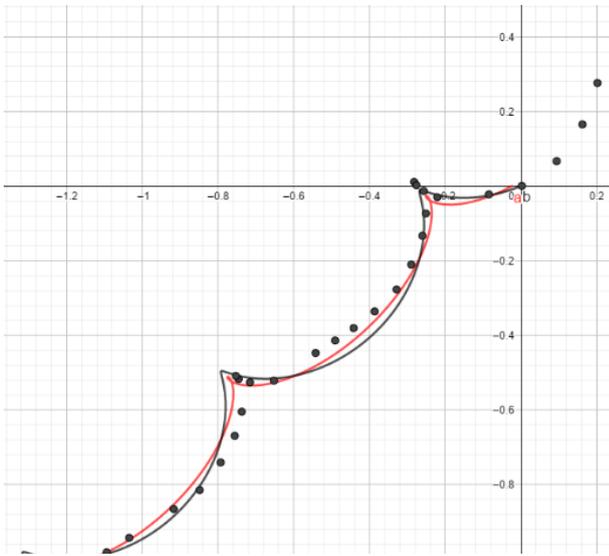

*Figure 14: Comparison of the measurements of a corner (points) with modelled curves (2a/b: black, 3a/b: red)*

6.4.2. VIDEO ANALYSIS OF THE FLIGHT OF A PAPER STRIP AT 7680FPS

From the summer onwards we used the Huawei P40 Pro camera. This smartphone is capable of recording at 7680 fps, which allowed us to view a smaller, more detailed section of the flight, thus giving us an even closer look at the flight trajectory.

We noticed that some strips bent a little as they fell. At the grammage of 205g/m² that we were mainly using, this primarily affected the very wide paper strips (>16 cm) of very limited height (< 3cm) and very great height (>14 cm). There were differences between the bends exhibited by these two types of strips, however. The very large strips changed their bending direction depending on the position of the strip: every time they reached a horizontal position, the normally vertical edges bent down slightly. This happened again after half a turn. The change in bending direction occurred when the strip was vertical [Fig. 14].

This led us to postulate certain assumptions: the reason for the bend is probably the greater drag at the centre of the strip. Here, the mass density is greater than at the edge, too, which favours this effect during rotation. In the position with the least drag, the deformation returns to its original state or even further because the mass of the strip is inert. This bending is reinforced by the fact that the drag in the ongoing rotation again has a more powerful effect on the strip. It then has the same pronounced bend as half a rotation previously, only in the other direction. By contrast, the direction of bending did not change for those paper strips that were of great width but limited height. No matter the position of the strip, the bend always remained the same. It seemed as if the vertical edges rotated with the centre around an axis running through two points of the strip. The narrow strips are much less stiff, which is why they bend much more easily. In the position with the least drag, there is no time to bend in the opposite direction due to the very high rotation frequency. So, the direction of the bend remains [Fig. 15].

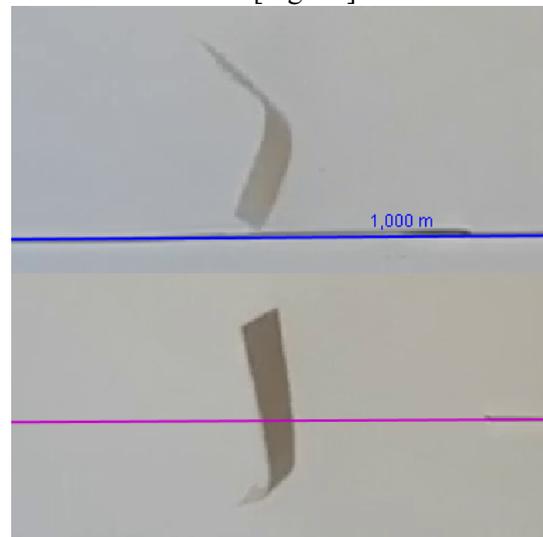

*Figure 15: bottom $t = t_o$, top $t = t_o + \frac{1}{2}\frac{1}{f}$ → bend is maintained*

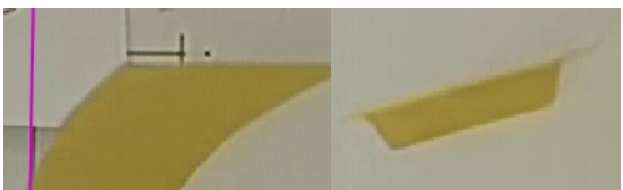

*Figure 13: left $t = t_o$, right $t = t_o + \frac{1}{2}\frac{1}{f}$ → bend changes*

## 6.5. FORCE ANALYSIS

By averaging different parameters, the physics of this movement can be described by means of classical mechanics [Fig. 16-18].

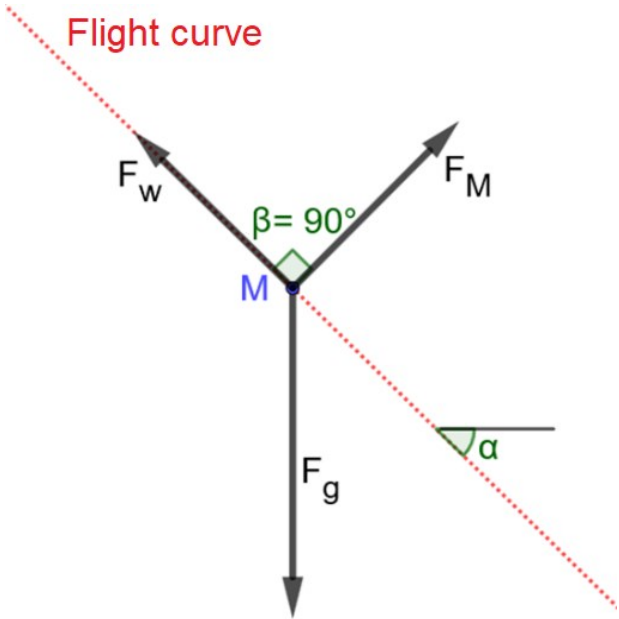

*Figure 18: Force analysis*

The following forces act on a strip of paper during flight

1. the weight force in the direction of the centre of the earth $F_G = m \cdot g$
2. the drag force against the direction of flight $F_w = \frac{\varrho}{2} \cdot A_p \cdot v^2 \cdot c_w$
3. the Magnus force orthogonal to the trajectory curve $F_m = \varrho \cdot v_s \cdot \omega \cdot h^2 \cdot b \cdot k$

Using a coordinate system on the trajectory curve, it is possible to obtain a force decomposition in two directions [Fig. 17].

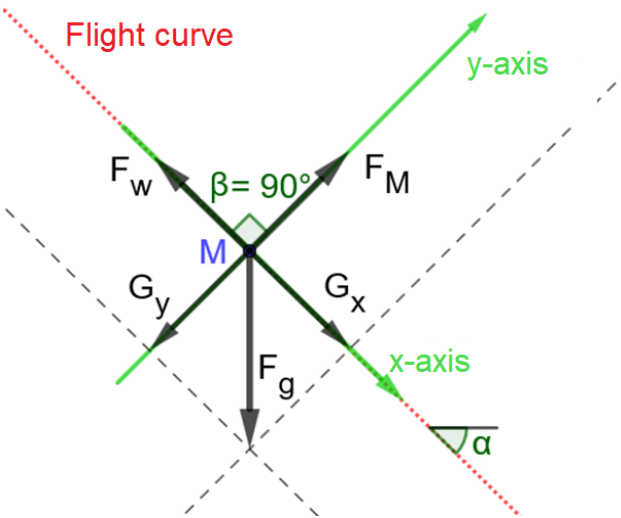

*Figure 17: Decomposition of forces*

In the y-direction: $\sum_{i=1}^{2} F_{iy} = F_m - G_y = 0$, as the two cancel each other out.

In the x-direction: $\sum_{i=1}^{2} F_{ix} = G_x - F_w = m \cdot a_x$, since in the x-direction the final resulting movement is that which has a force $F_{res}$ from $m \cdot a$ [Fig.18].

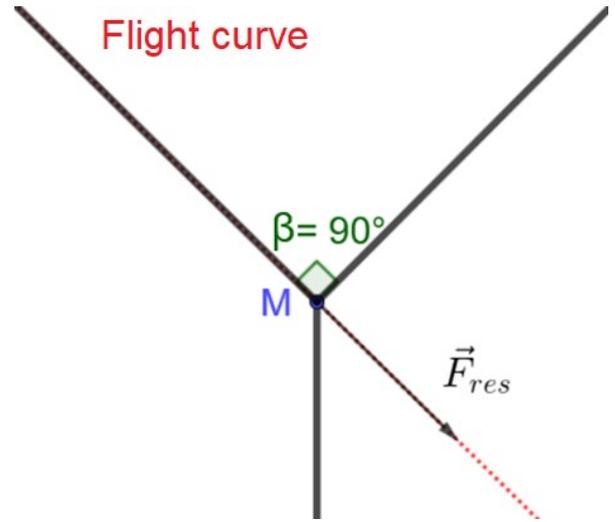

*Figure 16: Resulting force*

This means that for:
- $F_w$: $F_W = m \cdot (g \cdot sin(\alpha) - a)$, the following still applies: $F_w = \frac{\varrho}{2} \cdot A_p \cdot v^2 \cdot c_w$
- $F_m$: $F_m = m \cdot g \cdot cos(\alpha)$, the following still applies: $F_m = \varrho \cdot v_s \cdot \omega \cdot h^2 \cdot b \cdot k$

The following parameters are to be averaged:
- The cross-sectional area, as this is always different due to the rotation of the strip. The average cross-sectional area $A_p$ was taken into account by the factor $\frac{\sqrt{2}}{2} = sin(45°)$, by which the maximum area was multiplied: $A_p = \frac{\sqrt{2}}{2} \cdot A_{max}$
- The Magnus force: For the Magnus effect, a circle or, in 3D, a cylinder or sphere is generally assumed. If one assumes a complete revolution of the strip, this is geometrically similar in terms of fluid mechanics and we can approximately assume the lift effect. The approximation is taken into account by a correction factor k. [11][12]
- The velocity v, as this periodically increases and decreases due to rotation.
- The acceleration a, which is also not constant due to v.
- The angle α, as this can vary by a few degrees depending on the area of the trajectory curve.
- The frequency f, as we also take into account the median of this.
- The trajectory curve, as this is not a straight line but a rolling curve (see above).

For this decomposition of forces, both $m \cdot (g \cdot sin(\alpha) - a)$ and $\frac{\varrho}{2} \cdot A_p \cdot v^2 \cdot c_w$ therefore apply to $F_w$. $F_m$ is $m \cdot g \cdot cos(\alpha)$ and $\varrho \cdot v_s \cdot \omega \cdot h^2 \cdot b \cdot k$. Consequently, it is possible to determine the correction factors $c_w$ and k and then establish a trend for several of these.

The following applies for k: $k = \frac{m \cdot g \cdot cos(\alpha)}{\varrho \cdot v_s \cdot \omega \cdot h^2 \cdot b}$ and the following applies for $c_w$: $c_w = \frac{m \cdot (g \cdot sin(\alpha) - a)}{\frac{\varrho}{2} A_p \cdot v^2}$

We were able to obtain the angle of the flight path via the trajectory curve, the average speed via the v(t) diagram output by Tracker and the average acceleration via the derivative of v(t).

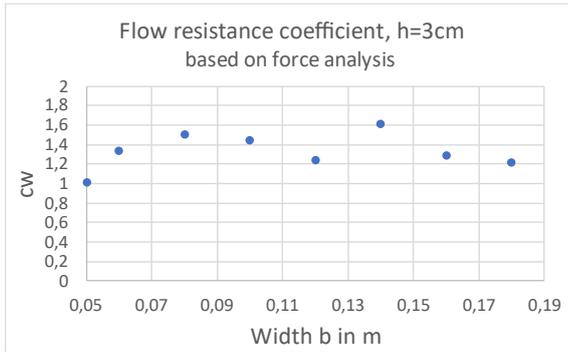

*Figure 20: Calculated cw value for h=3 cm*

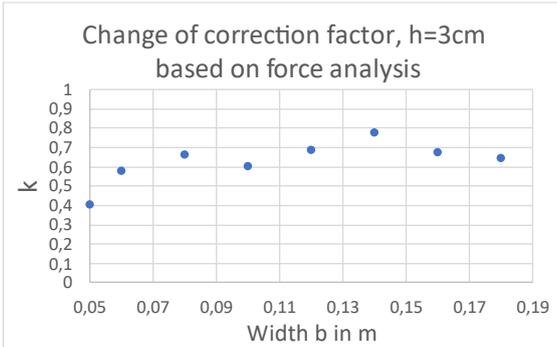

*Figure 21: Calculated correction factor for h=3 cm*

For the calculations shown above as examples (h=3 cm, G=205 g/m², b variable), the correction factors k scattered minimally between 0.4 and 0.8. The $c_w$ values for the same calculations are between 1 and 1.6 [Fig. 19, 20].

Since v, ω and α were the most error-prone, an error estimate was performed on these three parameters (see 6.4).

## 6.6. ENERGY ANALYSIS

In addition to analysing the forces, it is also possible to carry out an energy analysis.
For this we consider a rotation at a point in the flight where the frequencies of the rotation are constant. A rotation therefore takes Δt =1/f [in s]. In this time span, the strip has travelled the distance Δs = v·Δt, where v can be determined in Tracker. The velocity v can be assumed to be approximately constant during this period Δt. This means that the acceleration a is zero.
The height Δh that the strip has lost is Δh=Δs·sin(α), where α is the angle of the trajectory curve. This

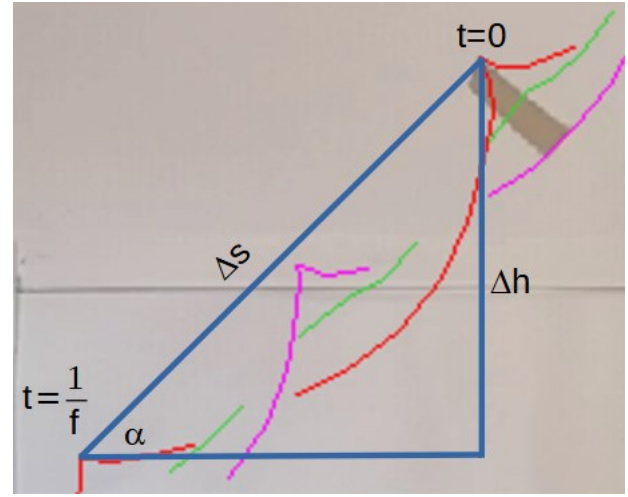

*Figure 19: Sketch for energy balance calculation*

angle was also determined using Tracker. [Fig. 21]
If we draw up an energy balance, ΔE = ΔE$_{pot}$ + ΔE$_{kin}$ + ΔE$_{rot}$ where ΔE$_{kin}$ = 0 and ΔE$_{rot}$ = 0, since v and f were assumed to be constant.
The following applies to the decrease in potential energy: ΔE$_{pot}$ = m·g·Δh
Therefore, ΔE = ΔE$_{pot}$ = m·g·Δh
The total loss of energy ΔE is converted into air turbulence (friction) in this approach.
For this reason, ΔE = F$_w$·Δs applies at the same time, where F$_w$ is the drag force for which the following applies: $F_w = \frac{\varrho}{2} \cdot A_p \cdot v^2 \cdot c_w$.
So m·g·Δh = $\frac{\varrho}{2} \cdot A_p \cdot v^2 \cdot c_w$ ·Δs
The drag coefficient $c_w$ can therefore be stated as
$c_w = \frac{m \cdot g \cdot \Delta h}{\frac{\varrho}{2} \cdot A_p \cdot v^2 \cdot \Delta s}$
If the result of the energy balance is compared to the result of the force calculation (see Tab. 3), we arrive at the following:

| Energy | Forces |
|---|---|
| $c_w = \frac{m \cdot g \cdot \Delta h}{\frac{\varrho}{2} \cdot A_p \cdot v^2 \cdot \Delta s}$ $= \frac{m \cdot g \cdot sin(\alpha)}{\frac{\varrho}{2} \cdot A_p \cdot v^2}$ | $c_w = \frac{m \cdot (g \cdot sin(\alpha) - a)}{\frac{\varrho}{2} \cdot A_p \cdot v^2}$ $= \frac{m \cdot g \cdot sin(\alpha)}{\frac{\varrho}{2} \cdot A_p \cdot v^2} - \frac{m \cdot a}{\frac{\varrho}{2} \cdot A_p \cdot v^2}$ |

*Table 2: Comparison of the calculation of the $c_w$ value based on an analysis of energy and forces*

The difference between the $c_w$ values calculated using the different considerations is therefore $\frac{m \cdot a}{\frac{\varrho}{2} \cdot A_p \cdot v^2}$.

Since a is very small for the force analysis, the $c_w$ values calculated in this way differ only in their order of magnitude of 0.4% (h=3 cm, b=18 cm, G=205 g/m²) to 5.6% (h=3 cm, b=6 cm, G=205 g/m²). Since the acceleration a decreased with increasing width, the difference between the $c_w$ value resulting from the force analysis and from the energy analysis decreases with increasing width. [Fig. 22]

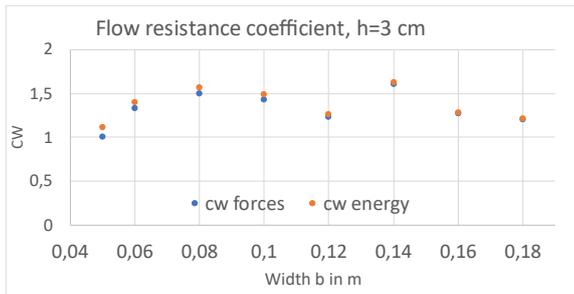

Figure 22: Comparison of the calculated cw values

## 7. ERROR ANALYSIS

### 7.1. DETECTING TRACKING ERRORS

Errors occur when setting the points on the corners, i.e. when tracking them: it often happened that a point was not placed precisely on the corner during fast tracking because the video resolution was too low. In addition, it sometimes happened that corners were mixed up during tracking because the camera perspective caused unavoidable optical illusions. As a result, orientation was sometimes lost. With Tracker, you can easily identify and correct these errors: as described above, the centroid is selected as a co-moving coordinate origin. For the points that did not match the anticipated pattern, it could be assumed that there was a tracking error because the dimensions of the strip did not change. It was then possible to remove or correct these missing points.
In Figure 23, you can see that at about second 7.5 the corners were swapped because the regular pattern of the graph is interrupted. The erroneous value at second 1.4 occurred because Tracker was not able to calculate the centroid due to a missing value. At second 0.5 and 0.9 in Figure 24, the strip was positioned very unfavourably in the air, so the corner was not well detected and incorrectly tracked.

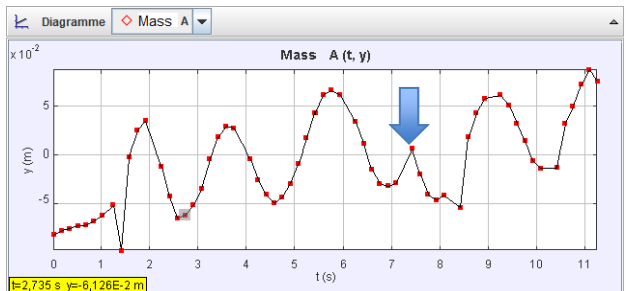

Figure 24: Tracking error (change of corner)

### 7.2. SYSTEM-RELATED TRACKING ERRORS

As in all experiments, errors can occur in tracking. In addition to reading errors for which we were responsible but which we were able to rectify, there were also errors in measurement accuracy that we were unable to actively address.
Since in our videos the paper strips always fell through the centre and this is hardly visible in low resolution, we can neglect the spherical aberration.
Before we used the Huawei P40 Pro, we had the problem of too few frames per second. When we calculated the rotation duration, the value per 360° rotation at constant rotation speed changed more often between two adjacent natural numbers. This problem results from insufficient video resolution. At 240fps, it often came down to half a frame. At a frequency of 25 Hz, one rotation takes 9.6 frames [Fig. 25], but it is only possible to measure 9 or 10 frames. The error then amounts to +-5%. This error does not appear in the median, only in the minima and maxima. The error can therefore be neglected for the purpose of calculation.

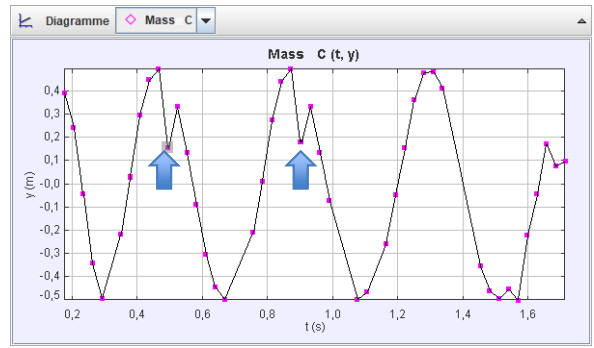

Figure 23: Tracking error (corner not positioned correctly)

| 1. Rotation | 2. Rotation | 3. Rotation | 4. Rotation | 5. Rotation | 6. Rotation |
|---|---|---|---|---|---|
| 0,104166667 | 0,1 | 0,095833333 | 0,1 | 0,104166667 | 0,1 |
| 0,1125 | 0,104166667 | 0,1 | 0,104166667 | 0,1 | 0,104166667 |
| 0,108333333 | 0,1 | 0,1 | 0,095833333 | 0,1 | 0,104166667 |
| 0,104166667 | 0,1 | 0,1 | 0,1 | 0,104166667 | 0,1 |

Figure 25: Rotation speed at insufficient video resolution

### 7.3. DISCARDED STRIPS

Our experimental objects were numerous and diverse. Our experiments only yielded good results when h/b was less than 1. Where h/b was almost 1, the strips demonstrated an unstable fall behaviour: in addition to normal rotation, the axis of rotation of the strips fluctuated back and forth. These experiments were therefore not suitable for analysis. With an h/b ratio of less than 1/10, the flight behaviour of the strip did not match that of most others: the axis of rotation of this very wide strip of limited height was now no longer in the surface area of the paper but only passed through two oscillating nodes of the strip. The vertical strip edges and the centre of the strip rotated around this axis. In the case of the strips with excessive mass per area, there were often only a few measurable rotations along the drop trajectory, even though we made full use of the room height. In order to be able to use these strips, the drop height would have to be increased even further. Our results therefore only apply to the value ranges of h, b and m within which we conducted our experiments. Nonetheless, it cannot be ruled out that the results might also apply to other value ranges such as those described above for the *Jufo* banner.

### 7.4. FORCE-ENERGY ERROR ANALYSIS

For the error analysis as applied to the force and energy breakdown, an error estimation for $c_w$ and k can be carried out using the linear error propagation law. For the purpose of error estimation, the fall angle $\alpha$, the velocity v, the acceleration a and the angular velocity ω are considered to be subject to error. The other variables are hardly subject to error and can be neglected here. Using partial derivatives, it is possible to determine the absolute error for $c_w$ and k.

The following therefore applies for energy: $\Delta c_{w1} = \left|\frac{\partial c_w}{\partial \alpha}\right| \cdot |\Delta\alpha| + \left|\frac{\partial c_w}{\partial v}\right| \cdot |\Delta v|$ with $c_w(\alpha, v) = \frac{m \cdot g \cdot \sin(\alpha)}{\frac{\varrho}{2} \cdot A_p \cdot v^2}$

or for the forces $\Delta c_{w2} = \left|\frac{\partial c_w}{\partial \alpha}\right| \cdot |\Delta\alpha| + \left|\frac{\partial c_w}{\partial v}\right| \cdot |\Delta v| + \left|\frac{\partial c_w}{\partial a}\right| \cdot |\Delta a|$ with $c_w(\alpha, v, a) = \frac{m \cdot g \cdot \sin(\alpha)}{\frac{\varrho}{2} \cdot A_p \cdot v^2} - \frac{m \cdot a}{\frac{\varrho}{2} \cdot A_p \cdot v^2}$

$\Delta c_{w1} = \left|\frac{m \cdot g \cdot \cos(\alpha)}{\frac{\varrho}{2} \cdot A_p \cdot v^2}\right| \cdot |\Delta\alpha| + \left|\frac{-2 \cdot m \cdot g \cdot \sin(\alpha)}{\frac{\varrho}{2} \cdot A_p \cdot v^3}\right| \cdot |\Delta v|$,

$\Delta c_{w2} = \left|\frac{m \cdot g \cdot \cos(\alpha)}{\frac{\varrho}{2} \cdot A_p \cdot v^2}\right| \cdot |\Delta\alpha| + \left|\frac{-2 \cdot m \cdot g \cdot \sin(\alpha)}{\frac{\varrho}{2} \cdot A_p \cdot v^3}\right| \cdot |\Delta v| + \left|\frac{m}{\frac{\varrho}{2} \cdot A_p \cdot v^2}\right| \cdot |\Delta a|$

For $\Delta\alpha$, $\Delta v$, $\Delta a$, $\Delta\omega$ a deviation of 5% from the original value is assumed.

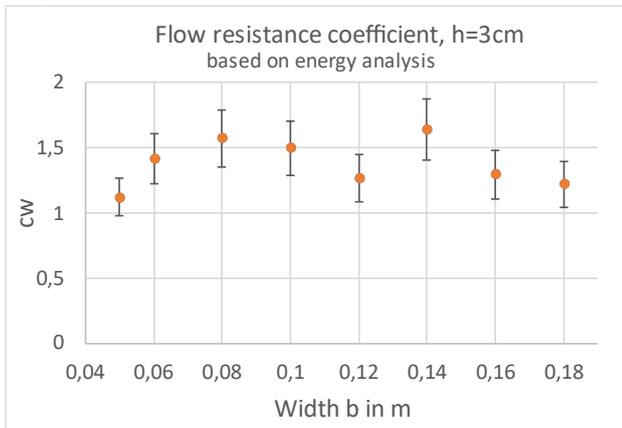

*Figure 27: Error in the energy analysis*

The error calculation was carried out for eight strips with a height of 3 cm and a width of 5 cm to 18 cm.

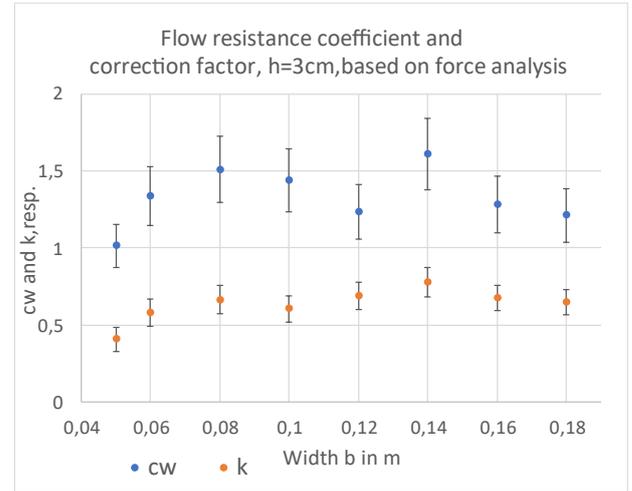

*Figure 26: Error in the force analysis*

The relative error $\frac{\Delta c_{w1}}{c_w}$ for the energy analysis is between 13% and 14.3% [Fig. 26], whereby the share of v in the error is about 2-3 times that of $\alpha$. The relative error $\frac{\Delta c_{w2}}{c_w}$ for the force analysis is between 13.9% and 14.5%, whereby the share of v in the error is about 2-3 times that of $\alpha$. Acceleration a hardly contributes to error (0.1% - 3.7%) The relative error $\frac{\Delta k}{k}$ for the correction factor in relation to the force analysis is between 12% and 19%, with the share of v and ω being equal and between 27% and 40%. [Fig. 27]

The error indicators for $c_w$ are in the same range as for k. Acceleration is not a factor here either. So for $c_w$ and k, a constant value can be assumed where relevant.

Another error analysis involving manual variation of the above-mentioned parameters resulted in a similar error spectrum.

### 8. DISCUSSION OF RESULTS AND SUMMARY

All in all, we can say that our project was successful. With well over 900 videos now recorded and analysed, we are able to make valid statements about the fall of paper strips for each of the 80 scenarios, and presumably for others, too.

In addition to accurately modelling the trajectory using cycloids, we found out how fast paper strips rotate depending on their dimensions and mass. For this purpose, we established a formula that allowed us to determine the rotation frequency of any strips within our measurement range. Furthermore, we can now say which forces act on the rotating strip and how strong they are. By way of conclusion, the fact that the determination of the $c_w$ value based on the force analysis and energy analysis provides equally significant values confirms that the chosen fluid mechanics assumption is approximately correct.

By carrying out an error analysis, we are able to specify a range for all our results within which our own measurements and any future measurements lie. Our results can be applied to geometries and grammages within our measuring range and to some extent to those outside this range.

It is worth noting that our project was based on very simple laws of physics. Although we initially thought the project was too complex, this assumption was belied by the fact that we unexpectedly obtained very clear results. The flight behaviour we observed can also be seen in larger rectangular objects, such as the base plates of the walkway to the platform above the Arecibo telescope when it collapsed [21] or the tumbling toast that was awarded the Ig Nobel Prize [22]. Confetti (in so-called "chaff") was used in World War II and is still used today in earth atmosphere research for radar deception as well as in the measurement of air currents [23]. So our research does have serious applications.

## 9. LATEST RESULTS

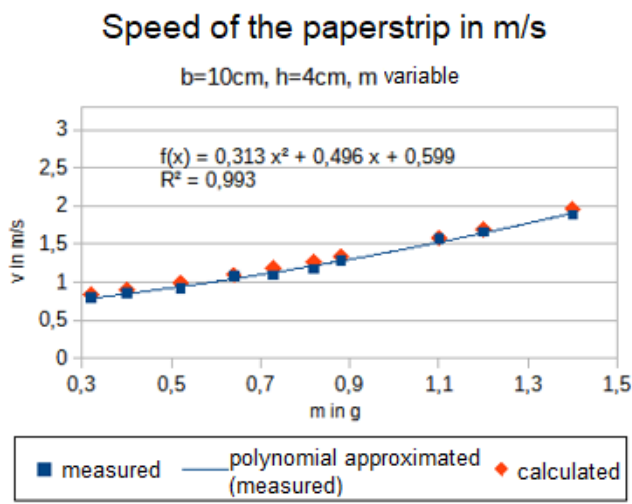

*Figure 29: Speed at variable height and constant width and mass*

Since completing the project in January, we have recorded almost 200 more videos to investigate flight distance, fall angle and fall speed in more detail. In this way we were able to establish promising correlations (within our research range) between the angle or speed of fall and the width, height and mass of the strips [Fig. 28]. Two comprehensive formulas such as the one for the frequency were obtained.

Finally, we performed an experiment to make the loss of energy visible. To do this, we let the paper strip fall through fog from a fog machine. In this way, we were able to visualize where the strip does work in the environment and confirm our considerations from the energy analysis [Fig. 29].

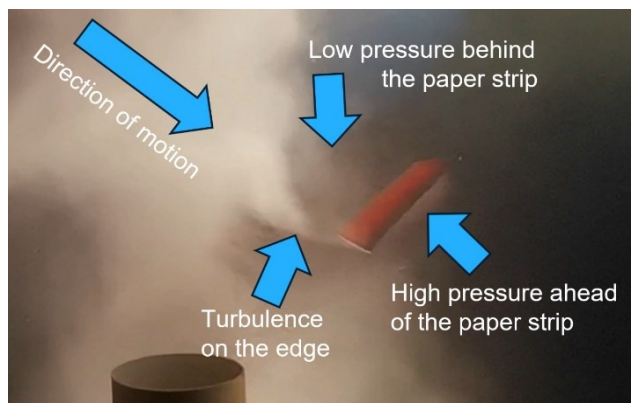

*Figure 28: Flying paper strips in fog*

## 10. OUTREACH

Despite our far-reaching investigations, the project still offers further potential for research:

So far, we have dealt exclusively with rectangular paper strips. Strips with "holes" in the surface or at the edge or other symmetrical paper shapes, such as ovals or circles, demonstrate the same flight behaviour. It may be possible to extend our rotation frequency formula here by a correction factor, e.g. for the effective area.

Our paper strips stayed in the air for a very long time as compared to other falling objects because of the drag. This is also something that is found in nature – especially with seeds: the most widely known example here is the maple with its "wings"; other examples are the hornbeam and the lime tree. Here again seeds rotate, so more time passes before they land on the ground and they are carried away by the wind. The interesting point here in contrast to our project is that maple seeds have a vertical axis of rotation. It would be fascinating to explore this similar yet very different type of "drag use". This is an area in which we have also carried out some preliminary tests. Research in this direction would result in an entirely new project, however.

## 11. List of sources and literature

## 12. Special software used

Tracker: https://physlets.org/tracker/
PyCharm: https://www.python.org/downloads/Pycharm
Geogebra: https://www.geogebra.org/
Solidworks: https://www.solidworks.com/de
Vernier Physics (iPad): https://www.vernier.com/product/video-physics-for-ios/